\documentclass[letterpaper, 11 pt]{article}  

\usepackage{graphics}           
\usepackage{epsfig}             
\usepackage{amsmath}            
\usepackage{amssymb}            
\usepackage{amsfonts}           
\usepackage{mathrsfs}           
\usepackage{hyperref}           
\usepackage[noabbrev]{cleveref} 
\usepackage{placeins}           
\usepackage{pgfplots}           
\usepackage{tikz}               
\usepackage{epstopdf}           
\usepackage{siunitx}            
\usepackage{empheq}             
\usepackage{booktabs}           
\usepackage{multirow}           

\newcommand{\setNumReal}{\ensuremath{\mathbb{R}}}

\newcommand{\mpi}[1][]{\ensuremath{{\pi}_{#1}}}
\newcommand{\mpnd}{\ensuremath{\mathcal{\#}}}

\renewcommand{\div}{\ensuremath{\mathrm{div}}}

\newcommand{\uvec}[1][]{\ensuremath{\mathbf{u_{#1}}(t)}}
\newcommand{\xvec}[1][]{\ensuremath{\mathbf{x_{#1}}(t)}}
\newcommand{\yvec}[1][]{\ensuremath{\mathbf{y_{#1}}(t)}}

\newcommand{\cvec}[1][]{\ensuremath{\mathbf{c_{#1}}(t)}}

\newcommand{\dxvec}[1][]{\ensuremath{\mathbf{\dot{x}_{#1}}(t)}}

\newcommand{\oxvec}[1][]{\ensuremath{\mathbf{x_{#1}}(0)}}

\newcommand{\txvec}[1][]{\ensuremath{\mathbf{x_{#1}}(T)}}
\newcommand{\tyvec}[1][]{\ensuremath{\mathbf{y_{#1}}(T)}}

\newcommand{\fyvec}[1][]{\ensuremath{\mathbf{y_{#1}}}}

\newcommand{\fvec}[1][]{\ensuremath{\mathbf{f_{#1}}}}

\newcommand{\abs}[1]{\lvert#1\rvert}
\newcommand{\norm}[1][]{\ensuremath{\left\lVert #1 \right\rVert}}

\newcommand{\sysA}[1][]{\ensuremath{A_\mathrm{#1}}}
\newcommand{\sysB}[1][]{\ensuremath{B_\mathrm{#1}}}
\newcommand{\sysC}[1][]{\ensuremath{C_\mathrm{#1}}}

\newcommand{\refA}{\ensuremath{A_\mathrm{r}}}
\newcommand{\refB}{\ensuremath{B_\mathrm{r}}}

\newcommand{\dela}{\ensuremath{\delta_\mathrm{a}(t)}}

\newcommand{\delr}{\ensuremath{\delta_\mathrm{r}(t)}}

\newcommand{\fb}{\ensuremath{K_1}}
\newcommand{\ff}{\ensuremath{K_2}}

\newcommand{\wgt}{\ensuremath{\mathbf{\hat{W}}(t)}}
\newcommand{\wgtT}{\ensuremath{\mathbf{\hat{W}}^T(t)}}

\newcommand{\dwgt}{\ensuremath{\mathbf{\dot{\hat{W}}}(t)}}
\newcommand{\owgt}{\ensuremath{\mathbf{{\hat{W}}}(0)}}
\newcommand{\evec}{\ensuremath{\mathbf{e}(t)}}
\newcommand{\evecT}{\ensuremath{\mathbf{e}^T(t)}}

\newcommand{\phix}[1][]{\ensuremath{\Phi\bigl(\mathbf{x_{#1}} (t)\bigr)}}

\graphicspath{{./figs/}}

\usetikzlibrary{plotmarks}
\usetikzlibrary{automata,arrows,positioning,patterns,decorations.pathmorphing,calc}
\usetikzlibrary{shapes}

\newlength\figureheight
\newlength\figurewidth
\crefname{figure}{Fig.}{Figs.}
\Crefname{figure}{Fig.}{Figs.}
\crefname{table}{Table}{Tables}
\Crefname{table}{Table}{Tables}

\crefformat{equation}{(#2#1#3)}
\crefrangeformat{equation}{(#3#1#4) to~(#5#2#6)}
\crefmultiformat{equation}{(#2#1#3)}%
{ and~(#2#1#3)}{, (#2#1#3)}{ and~(#2#1#3)}

\newcommand{\footremember}[2]{%
\footnote{#2}
\newcounter{#1}
\setcounter{#1}{\value{footnote}}%
}
\newcommand{\footrecall}[1]{%
\footnotemark[\value{#1}]%
}

\title{\LARGE \bf
Measures and LMIs for Lateral F-16 MRAC Validation
}

\author{Daniel Wagner\footremember{cvut}{D. Wagner, D. Henrion, and M. Hrom\v{c}\'{i}k are with the Faculty of Electrical Engineering, Czech Technical University in Prague, Technick\'{a} 2, CZ-16626 Prague, Czech Republic {\tt\small \{wagneda1, henridid, hromcik@fel.cvut.cz\}}}%
, and Didier Henrion\footrecall{cvut} \footnote{D. Henrion is with CNRS, LAAS, 7 avenue du colonel Roche, F-31400 Toulouse, France {\tt\small henrion@laas.fr}}
, and Martin Hrom\v{c}\'{i}k\footrecall{cvut}
\footnote{This work is supported by the Czech Science Foundation (GA\v{C}R) under contract number GA16-19526S}
}

\begin{document}

\maketitle

\begin{abstract}
Occupation measures and linear matrix inequality (LMI) relaxations (called the moment sums of squares or Lasserre hierarchy) are state-of-the-art methods for verification and validation (VV) in aerospace. In this document, we extend these results to a full F-16 closed-loop nonlinear dutch roll polynomial model complete with model reference adaptive control (MRAC). This is done through a new technique of approximating the reference trajectory by exploiting sparse ordinary differential equations (ODEs) with parsimony. The VV problem is then solved directly using moment LMI relaxations and off-the-shelf-software. The main results are then compared to their numerical counterparts obtained using traditional Monte-Carlo simulations.
\end{abstract}


\section{INTRODUCTION}
\label{sec:acc_intro}

Model reference adaptive control (MRAC) has been researched extensively by the aerospace community in the last few decades. Examples include the successful flight testing the X-36 Tailless fighter \cite{ref:x36} and the JDAM guided munitions \cite{ref:jdam}. One of the main benefits of adaptive controllers is their capability of handling adverse conditions and/or inherent uncertainty in the aircraft dynamics. Despite its advantages, there are some entry barriers for practical use of MRAC:
\begin{enumerate}
  \item The closed-loop stability relies on its Lyapunov candidate function. An engineer wishing to use MRAC must have a solid theoretical background in nonlinear control. For non-autonomous systems, the best that can be achieved theoretically with this framework is asymptotic stability. 
  \item If any assumptions for the Lyapunov candidate function are violated (for example, when there are unstable reference trajectories \cite{ref:johnson}), then instability in the closed-loop model may exist. From a safety-critical standpoint, this is undesirable. 
\end{enumerate}
Consequently, there exists no formal procedure by the Federal Aviation Administration (FAA) for VV of MRAC for national air and space \cite{ref:closing}. 

One possible solution is Monte-Carlo, which is already widely used in VV because it is very robust. This approach becomes computationally difficult when there are a large number of uncertainties in the state-space. Conversely, if the state-space is not explored adequately, the Monte-Carlo framework may not even detect unsafe trajectories. This problem is exacerbated when trying to validate closed-loop models that use sophisticated feedback laws such as MRAC. Challenges with using Monte-Carlo to certify MRAC are discussed extensively in \cite{ref:roadmap} and \cite{ref:closing}. 

A recent development in the verification and validation (VV) for aerospace is using of moment sum of squares (SOS) with off-the-shelf-software. The authors of \cite{ref:robust,ref:balas} focus on polynomial dynamical models and polynomial SOS Lyapunov candidate functions. This VV methodology was also used for assessing robust stability of space launcher control laws within the SAFE-V project \cite{ref:spacelauncher}. In \cite{ref:cdc}, this framework was used to certify closed-loop models with MRAC. 

Like in our previous work \cite{ref:cdc}, we wish to close the numerical gap by validating closed-loop models with MRAC controllers using our VV framework. We also use new theoretical framework provided by \cite{ref:matteo} where the complexity of solving the LMI relaxations is reduced by exploiting sparsiry of the ordinatry differential equations (ODEs). {In particular, we are interested in qualitative properties such as safety (all trajectories starting from a set of initial conditions never reach a set of bad states), avoidance (at least one trajectory starting from initial conditions will never reach a set of bad states), eventuality (at least one trajectory starting from a set of initial conditions will reach a set of good states in finite time), reachability (at least one trajectory starting from a set of initial conditions will reach a set of good states in finite time), and robustness (all trajectories from a set of initial conditions guarantee acceptable performance subject to disturbances and/or unmodeled dynamics).}

{The procedure follows directly from \cite{ref:spacelauncher}, see also \cite{ref:Henrion} for a broader perspective. We first rephrase our validation problem as a robustness analysis problem and then as a nonconvex nonlinear optimization problem over admissible trajectories. Then the problem is expressed equivalently as an infinite dimensional linear programming (LP) problem by introducing occupation measures supported over admissible trajectories. We finally relax the infinite dimensional LP problem of measures to a finite dimensional linear matrix inequality (LMI) problem of moments. The solutions to our VV problem are primal in the sense that we optimize directly over the system trajectories. The well-established Lyapunov certificates can also be retrieved from the dual SOS LP problem.
Off-the-shelf-software (Gloptipoly 3 for MATLAB\cite{ref:gloptipoly}) and SDP solvers (such as MOSEK \cite{ref:mosek} or SeDuMi \cite{ref:sedumi}) can be used with our framework.}

{The main contributions are as follows:
\begin{itemize}
\item We describe a nonlinear dutch-roll F-16 polynomial model that we want to validate. A closed-loop model is derived using the standard LQR + MRAC augmentation. The controller is simplified by building the adaptive feedback around the aileron channel. 
\item For the LQR + MRAC augmentation problem is solved by exploiting sparsity \cite{ref:matteo} for ODEs. This is done by approximating the reference trajectories with a given roll angle command signal. This procedure reduces the size of the largest SDP block. Compared to \cite{ref:cdc}, we more than double the amount of states (from 4 states to 9 total) we can validate. 
\item Our VV framework is used to certify the closed-loop model subject to normal and reduced control effectiveness. We compare the baseline LQR closed-loop model to the same model using the LQR + MRAC augmentation. Likewise, we certify the same closed-loop model using Monte-Carlo methods. We can guarantee convergence of our trajectories in finite time, unlike the results provided in \cite{ref:robust} and \cite{ref:balas}. This is better than Lyapunov certificates that use Barbalat's lemma \cite{ref:14}, which only guarantees convergence in infinite time. 
\end{itemize}  }

The organization of this document is as follows: \Cref{sec:acc_dutch} discusses the nonlinear closed-loop F-16 polynomial model and the MRAC augmentation, \Cref{sec:acc_main} contains our main numerical results, and \Cref{sec:acc_conc} contains a small discussion of our conclusions and future results.


\section{F-16 DUTCH-ROLL DYNAMICS}
\label{sec:acc_dutch}
The trimmed nonlinear dutch-roll dynamics of an F-16 traveling at $502$ ft/s and $\alpha = 2.11 \ \mathrm{deg}$ is given by
\begin{equation}
  \begin{aligned}
    \dxvec[q] &= {\sysA} \xvec[q] + {\sysB} \Lambda \bigl(\uvec \\&\qquad + \Delta(\xvec[q],\uvec)\bigl), \quad \oxvec[q] = \mathbf{x_{q0}} \\
            \yvec &= {\sysC} \xvec[q],
  \end{aligned}
  \label{eqn:acc_lat}
\end{equation}
where 
\begin{gather}
\sysA = \begin{bsmallmatrix} -0.3220 & 0.0640 & 0.0364 & -0.9917 \\
             0 & 0 & 1 & 0.0393 \\ 
             -30.6490 & 0 & -3.6784 & -0.6646 \\
            8.3595  & 0 & -0.0254 & -0.4764\end{bsmallmatrix}, \\ \sysB = \begin{bsmallmatrix} 0 & 0 \\ 0 & 0 \\ 
            -0.7331 & 0.1315 \\ 
            -0.0319 & -0.0620\end{bsmallmatrix}, \quad \sysC = \begin{bsmallmatrix} 1 & 0 & 0 & 0 \\
             0 & 1 & 0 & 0 \end{bsmallmatrix},
\end{gather}
 $\xvec[q] = \begin{bsmallmatrix} \beta(t) & \phi(t) & p(t) & r(t) \end{bsmallmatrix} \in \mathbb{R}^4$ is the lateral state vector, $\uvec = \begin{bsmallmatrix}\dela & \delr\end{bsmallmatrix} \in \mathbb{R}^2$ are the measurable control inputs, $\yvec \in \mathbb{R}^2$ is the output, {$\Lambda = \lambda I_{2 \times 2}$, $\lambda \in \mathbb{R}_+$} is the control effectiveness, and $\Delta(\xvec[q],\uvec) \in \mathbb{R}^2[x]$, {found below in \cref{eqn:acc_Delta}}, contains unknown higher order dynamics which include dead-zone and loss of control effectiveness for large $\beta(t)$. The open-loop flight model, coefficients, and the nonlinearities were taken from \cite{ref:lavretsky}. The higher order terms in $\Delta(\xvec[q],\uvec)$ were derived by taking the Taylor series of hyperbolic functions. In the proceeding subsection we discuss the derivation of the nominal and adaptive control law used in the closed-loop model.
\begin{figure*}[tb]
\begin{equation}
\begin{aligned}
\Delta(\xvec[q],\uvec) &=
\left[\begin{smallmatrix}
  (1-\num{4.2646} \beta^2(t)) (\num{9.0028E-7} \dela - \num{6.0019E-7} \dela + \num{0.001} \dela) \\
  (1-\num{4.2646} \beta^2(t)) (\num{3.6317E-04} \delr + \num{2.4205E-04} \delr + \num{0.001} \delr)
\end{smallmatrix}\right.\\
&\quad\quad
\left.\begin{smallmatrix}
  +  \num{0.0750} (-\num{0.125}+\num{0.07854} p(t) - \num{0.0013708} p^2(t)) \\
  +  \num{0.4500} (-\num{0.125}+\num{0.07854} p(t) - \num{0.0013708} p^2(t))
\end{smallmatrix}\right.\\
&\qquad\qquad
\left.\begin{smallmatrix}
  \cdot(\num{0.05236}r(t) + 1) \\
  \cdot(\num{0.05236}r(t) + 1) 
\end{smallmatrix}\right] 
\end{aligned}
\label{eqn:acc_Delta} 
\end{equation}
\hrulefill
\end{figure*}


\subsection{Closed-Loop Configuration}
Consider the lateral dynamics in the form of \cref{eqn:acc_lat}. Our control objective is to asymptotically track the reference trajectory
\begin{equation}
  \begin{aligned}
    \dxvec[r] &= \refA \xvec[r] + \refB \cvec, \quad \oxvec[r] = \mathbf{x_{r0}} \\
    \yvec[r] &= C \xvec[r]
  \end{aligned}  
  \label{eqn:acc_ref}
\end{equation}
where $\cvec = \begin{bsmallmatrix} \beta_\mathrm{cmd} & \phi_\mathrm{cmd} \end{bsmallmatrix} \in \mathbb{R}^2$ are the given piecewise continuous bounded yaw/roll command signals, and $\xvec[r] \in \mathbb{R}^4$ is the reference state vector. Nominal controller gains 
\begin{gather*}  
  \fb = \begin{bsmallmatrix} 10.6901 & -9.5824 & -2.0328 & -6.1944 \\
                             -0.3982 & -0.2043 & -0.4170 & -27.0142  
        \end{bsmallmatrix}, \\
  \ff = \begin{bsmallmatrix} -2.9031 & -9.9924 \\ 156.5907 & -2.4300 \end{bsmallmatrix},
\end{gather*}
were derived using the LQR-method \cite{ref:lqr} such that $A_\mathrm{r} = A - B \fb$ is Hurwitz, $B_r = B \ff$, and the DC gain between the command signals $\cvec$ and output $\yvec$ is unity as $t \rightarrow \infty$. Now consider the combined nominal/adaptive feedback law 
\begin{equation}
\uvec = \uvec[n] + \uvec[a], 
\label{eqn:acc_u}
\end{equation}
with \emph{baseline} nominal control law $\uvec[n] = \fb \xvec[q] + \ff \cvec \in \mathbb{R}^2$ and adaptive control law $\uvec[a] = - \wgtT \phix$. {Basis function $\Phi_i (\xvec) = (1 + e^{x_{q,i}})^{-1}$, $i = 1,\dots,4$} is known and $\wgt \in \mathbb{R}^1$ satisfies the weight update law
\begin{multline}
  \dwgt =  \underbrace{\begin{bsmallmatrix} \epsilon & & & \\ & 300 & & \\ & & \epsilon & \\ & & & \epsilon \end{bsmallmatrix}}_\Gamma \phix[q] \evecT P \underbrace{\begin{bsmallmatrix} 0 & 0 \\ 0 & 0 \\ 
            -0.7331 & 0 \\ 
            -0.0319 & 0 \end{bsmallmatrix}}_{\sysB[ail]}, \\ \owgt = \mathbf{\hat{W}_0}
  \label{eqn:acc_wgt} 
\end{multline}
where $\epsilon << 1$, $\evec = \xvec[q] - \xvec[r]$, and positive definite symmetric $P \in \mathbb{R}^{4 \times 4}$ is the unique solution to the Lyapunov equation 
\begin{equation}
  0 = \refA^T P + P \refA + 100 I_{4 \times 4}. 
  \label{eqn:acc_lyap}
\end{equation}
Theorems that highlight the boundedness and performance of this configuration can be found in \cite{ref:mrac}.

{States containing small gain $\epsilon$ are removed in this configuration because they contribute little to the feedback}. There is only adaptive feedback through the aileron channel in \cref{eqn:acc_u,eqn:acc_wgt}. This is done to reduce the total number of states in the system. We will demonstrate the validity of our approach in the proceeding section. 

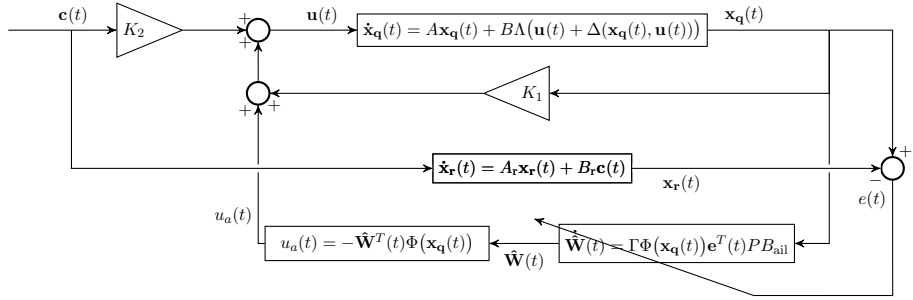
\begin{figure}[ht]
\centering
\tikzset{
	block/.style = {draw, rectangle, align = center},
	note/.style = {draw = red, thick, dotted, ellipse, align = center},
    sum/.style = {draw, thick, circle, inner sep = 0cm, minimum size = 0.5cm},
    tip/.style = {->, >=stealth'},
    redtip/.style = {->, >=stealth', red, dash dot},
    rtip/.style = {<-, >=stealth'},
    input/.style = {coordinate},
    doubletip/.style = {white, line width=3pt, line cap=butt},
    gain1/.style = {isosceles triangle, draw, node distance=1cm}, 
    gain2/.style = {isosceles triangle, draw, node distance=1cm, shape border rotate=180} 
}
\begin{tikzpicture}[ >=stealth, auto, node distance=1.4cm, scale=0.6, every node/.style={transform shape}]
    \node [input, name=input] {};
    \node [coordinate, right =of  input] (in) {};
    \node [gain1, right = of in] (k2) {$K_2$};
    \node [sum,  right = of k2] (u) {};
    \node [block, right =5 em of u, minimum height=0.7cm] (unc) {$\dxvec[q] = \sysA \xvec[q] + \sysB \Lambda \bigl(\uvec + \Delta(\xvec[q],\uvec)\bigl)$};
    \node [coordinate, right =7 em of unc] (pt1) {};
    \node [coordinate, right =of  pt1] (pt2) {}; 
    \node [coordinate, below =of  pt2] (pt3) {};
    \node [coordinate, left =of pt3] (pt4) {};
    \node [sum, below =of pt3] (sum1) {}; 
    \node [coordinate, below =of sum1] (pt5) {};
    \node [coordinate, below =3 em of pt5] (pt6) {}; 
    \node [gain2] (k1) at (unc |- pt4) {$K_1$};
    \node [sum] at (u |- k1) (sum2) {};
    \node [block]  at (unc |- sum1) (ref) {$\dxvec[r] = \refA \xvec[r] + \refB \cvec$};
    \node [coordinate, left =of pt5] (pt8) {};
    \node [block, left =2 em of pt8] (wgt) {$\dwgt = \Gamma \phix[q] \evecT P \sysB[ail]$}; 
    \node [coordinate, above left=2 em of wgt.180] (pt7) {};
	\node [coordinate, left =8 em of pt6] (pt9) {}; 
    \node [block, left =4 em of wgt] (adp) {\begin{tabular}{c} $u_a(t) = -\wgtT \phix[q]$ \end{tabular}};  
    \draw [tip] (unc) -| node[pos=0.1] {$\xvec[q]$} node [pos=0.98] {$+$} (sum1);
    \draw [tip] (k2) -- node [pos=0.98, above] {$+$} (u);
    \draw [tip] (pt1) |- (k1);
    \draw [tip] (u) -- node[pos=0.6] {$\uvec$}  (unc);
    \draw [tip] (sum1) -- node[pos=0.3, left] {$e(t)$} (pt5) -- (pt6) -- (pt9) -- (pt7);
    \draw [tip] (pt1) |- (wgt); 
   	\draw [tip] (k1) -- node [pos=0.98, below] {$+$}  (sum2);
    \draw [tip] (wgt) -- node {$\wgt$} (adp);
    \draw [tip] (adp) -| node[pos=0.6] {$u_a(t)$} node [pos=0.98] {$+$} (sum2); 
    \draw [tip] (sum2) -- node [pos=0.98] {$+$} (u); 
    \draw [doubletip] (in) |- (ref);
    \draw [tip] (input) -- node[pos=0.6] {$\cvec$}  (k2);
    \draw [tip] (in) |- (ref); 
    \draw [doubletip] (ref) -- (sum1);
    \draw [tip] (ref) -- node [pos=0.2,below] {$\xvec[r]$} node [pos=0.98, below] {$-$} (sum1);     
    \node [block] at (unc |- sum1)  (ref2) {$\dxvec[r] = \refA \xvec[r] + \refB \cvec$};
\end{tikzpicture}{}
\caption{Dutch-Roll Closed-Loop Configuration}
\label{fig:acc_block}
\end{figure}

We can now write the closed-loop model with \cref{eqn:acc_lat,eqn:acc_ref,eqn:acc_u,eqn:acc_wgt} in their compact form 
\begin{equation}
  \dxvec = \fvec\bigl(t,\xvec, \Lambda \uvec\bigl),
  \label{eqn:acc_fx}
\end{equation}
where $\xvec= \begin{bsmallmatrix} \xvec[q] & \wgt & \xvec[r] \end{bsmallmatrix} \in \mathbb{R}^{9}$. The unused rudder states in the weight update law \cref{eqn:acc_wgt} are discarded. A block diagram of the complete closed-loop model is provided in \cref{fig:acc_block}.


\section{MAIN NUMERICAL RESULTS}
\label{sec:acc_main}
We wish to validate our existing closed loop polynomial aircraft model \cref{eqn:acc_fx} by finding the initial state that maximizes of the norm of the concave cost function $J = - \norm[\cvec - \tyvec]^2_2$ with given terminal time $T = 10 \ \mathrm{s}$ and $\cvec = \begin{bsmallmatrix} 0 & +10\end{bsmallmatrix} \frac{\pi}{180}$. If we can show that for every chosen initial state
\begin{multline*}
\oxvec \in X_0 \triangleq \begin{bsmallmatrix} -10 & +10 \end{bsmallmatrix}^4 \frac{\pi}{180} \times \begin{bsmallmatrix} -0.001 & +0.001 \end{bsmallmatrix} \\ \qquad \times \begin{bsmallmatrix} -0.001 & +0.001 \end{bsmallmatrix}^4 \frac{\pi}{180}
\end{multline*}
that all trajectories remain bounded in
\[
\xvec\in X \triangleq \begin{bsmallmatrix} -30 & +30 \end{bsmallmatrix}^8 \frac{\pi}{180} \times \begin{bsmallmatrix} -80 & +80 \end{bsmallmatrix}
\]
until they reach the final state belonging to a set $\txvec \in \{ J \leq \num{0.003} \}$, then the control law is validated.

The main results of this section rely heavily on theoretical background discussed extensively in \cite{ref:Henrion,ref:cdc,ref:affine}. The procedure consists of writing our validation problem as a piecewise polynomial dynamical optimization problem 
\begin{equation}
  \begin{aligned}
  J = \ & \inf & & h_T (T, \xvec) + \int_0^T h(t,\xvec) dt & \\
  & \ \text{s.t.} & &  \dxvec = \fvec_j(t,\xvec), \quad j = 1,2,   &\\
    & & & \xvec\in X_j, \quad \oxvec \in X_0, \quad \txvec\in X_T, \quad t \in [0,T] &
  \end{aligned}
  \label{eqn:acc_polynomial}
\end{equation}
{with given polynomial dynamics $f \in \mathbb{R}[t,x]$ and costs $h, h_T \in \mathbb{R}[t,x]$, and state trajectories $\xvec$ constrained in the compact basic semialgebraic sets} $X$, $X_0$, and $X_T$.

We then write \cref{eqn:acc_polynomial} as its infinite-dimensional measure-LP problem {
\begin{equation}
  \begin{aligned}
  J_\infty = \ & \inf & & \int h_T(T,\xvec) d \mu_T + \int h(t,\xvec) d \mu & \\
  & \ \text{s.t.} & &   \frac{\partial \mu}{\partial t} + \div \fvec_j \mu_j + \mu_T = \mu_0 & \\
    & & & \int \mu_0 = 1 
  \end{aligned}
  \label{eqn:acc_measure} 
\end{equation}
where $\div $ is the divergence operator and the infimum is with respect to the occupation measure $\mu \in \mathscr{M}_+([0,T] \times X)$, initial measure $\mu_0 \in \mathscr{M}_+(\{0\} \times X_0)$, terminal measure $\mu_T \in \mathscr{M}_+(\{T\} \times X_T)$, and terminal time $T > 0$.} This infinite dimensional problem of measures can be relaxed to a finite moment LMI problem of truncated sequences sequences, using Lasserre's LMI hierarchy \cite{ref:lassere}. When relaxation order $d \in \mathbb{N}$ tends to infinity, it holds that $J_d \leq J_{d+1} \leq J_\infty$ and $\lim_{d \rightarrow \infty} J_d = J_\infty$. 

A piecewise disturbance is also included where the aircraft experiences reduced control effectiveness at large roll angles. To include the disturbance, we reformulate the optimization problem
with the system dynamics defined as locally affine functions in two cells $X_j$, $j = 1,2$ corresponding respectively to the regimes of the disturbance
\begin{equation}
  \begin{aligned}
      X_1 &\triangleq \{\xvec \in \mathbb{R}^7: \abs{\phi} \leq \phi_\mathrm{max} \}, \quad \lambda = 1 \\
      X_2 &\triangleq \{\xvec \in \mathbb{R}^7: \abs{\phi} \geq \phi_\mathrm{max}\}, \quad \lambda = 0.2
  \end{aligned}
\end{equation}
such that $\phi_\mathrm{max} \in \mathbb{R}_+$. 

For our main results, we consider two cases: $\phi_\mathrm{max}  = 1$ and ${\phi_\mathrm{max} = \num{0.314159}}$. We first consider \cref{eqn:acc_fx} without the MRAC augmentation ($\uvec[a] = 0$) for both cases.  Then in \cref{sec:acc_pars} we discuss our main contribution for implementing \cref{eqn:acc_fx} with the MRAC augmentation by reducing the total number of states using parsimony. Lastly, the results for \cref{eqn:acc_fx} with adaptive feedback are presented in \cref{sec:acc_mrac} and compared against the baseline control law from \cref{sec:acc_lqr} for both cases. 

If \cref{eqn:acc_polynomial} can be written in the form of \cref{eqn:acc_measure}, it can be solved directly using off-the-shelf-software. {In our examples, we used GloptiPoly 3 \cite{ref:gloptipoly} and MOSEK \cite{ref:mosek} to solve a hierarchy of moment LMI relaxations.} Additional steps must be taken to reduce numerical problems. In \cref{eqn:acc_fx}, we employ a normalizing matrix $D = \mathrm{diag}[{a_1, \dots, a_9}]$, $a_1,\dots,a_9 \in \mathbb{R}_+$ such that
\begin{equation}
  \dxvec = T D \fvec_j\bigl(t, D^{-1} \xvec, \Lambda \uvec \bigl).
  \label{eqn:acc_cfx}
\end{equation}
and all trajectories, including the time domain, are normalized  within the interval $\begin{bsmallmatrix}-1, & +1\end{bsmallmatrix}$.  
 
Our numerical results can all be found in \cref{tab:phimax1,tab:phimax2}. Then the results obtained with our framework are compared against Monte-Carlo. The Monte-Carlo simulations in \cref{fig:acc_phimax1,fig:acc_phimax2} were built by using Newton's Method ($t_\mathrm{step} = 0.001 \ \mathrm{s}$) and nested for-loops with evenly spaced initial conditions. The green lines denote the desired closed-loop performance. The numerical maximum upper bounds found in \cref{tab:acc_mc} were obtained by searching for the largest $J$ generated by every initial condition. 

\subsection{Baseline Controller Problem}
\label{sec:acc_lqr}

The baseline controller configuration can be obtained by setting $\Gamma = 0$ and $\uvec[a] = 0$. The closed-loop model does not depend on the error dynamics $\evec$ in this form, so we also negate the reference dynamics \cref{eqn:acc_ref}. Given \cref{eqn:acc_fx}, the polynomial dynamical optimization problem becomes 
\begin{equation}
  \begin{aligned}
  J = \ & \inf_{\yvec} & & - \norm[\cvec - \tyvec]^2_2 & \\
  & \ \text{s.t.} & &  \dxvec = T D \fvec_j \bigl(t, D^{-1} \xvec, \Lambda_j \uvec \bigl)  &\\
  & & &  \xvec\in X_j, \quad t \in [0,1], \quad j=1,2 &\\
    & & & \oxvec \in X_0, \quad \txvec\in X_T, &
  \end{aligned}
  \label{eqn:acc_poly_lqr}
\end{equation}
and its measure LP 
\begin{equation}
  \begin{aligned}
  J_\infty = \ & \inf & & -\int \norm[\cvec - \tyvec]^2_2 \mu_T & \\
  & \ \text{s.t.} & &  \frac{\partial \mu_j}{\partial t} + \div \fvec_j \mu_j + \mu_T = \mu_0 & \\
    & & & \int \mu_0 = 1. 
  \end{aligned}
  \label{eqn:acc_meas_lqr}
\end{equation}
The moment LMI relaxations problem can now be obtained directly from \cref{eqn:acc_meas_lqr}. 

As shown in the Monte-Carlo simulations \cref{fig:acc_phimax1,fig:acc_phimax2,tab:acc_mc}, there is no degradation in tracking performance when ${\phi_\mathrm{max} = 1}$. When ${\phi_\mathrm{max} = \num{0.314159}}$, trajectory overshoot is heavily penalized. Only $2\%$ of the trajectories fail to achieve proper tracking performance in \cref{fig:acc_phimax2}. Insufficient sampling of the state-space in the Monte-Carlo simulations could result these trajectories remaining undetected. 

On the other hand, our framework can extract directly the unsafe trajectories. \Cref{tab:phimax1,tab:phimax2} contains the Gloptipoly 3 + MOSEK results for both cases. For ${\phi_\mathrm{max} = 1}$, the baseline controller achieves our desired terminal cost. This is not reflected for ${\phi_\mathrm{max} = \num{0.314159}}$, which produces a significantly larger upper bound. This indicates a loss in tracking performance.  

\FloatBarrier
{\subsection{Exploiting Sparsity for ODEs}}
\label{sec:acc_pars}
 
{Consider the polynomial dynamics in the form of
 \begin{equation}
 \begin{aligned}
     \dxvec[1] &= \fvec[1]\bigl(t,\xvec[1],\yvec\bigl)   \\
     \dxvec[2] &= \fvec[2]\bigl(t,\xvec[2]\bigl), \quad \yvec = \sysC[2] \xvec[2]
  \end{aligned}
 \label{eqn:acc_pfx}
 \end{equation}
 where $\xvec[1], \ \xvec[2] \in \setNumReal^n$, $\yvec \in \setNumReal^m$, and $m < n$. The dynamics of $\fvec[2](\cdot)$ are \emph{autonomous} and serves as a control input for $\fvec[1](\cdot)$.} In this configuration, \cref{eqn:acc_pfx} can be \emph{approximated} using the sparse measure LP
\begin{equation}  
  \begin{aligned}
  J_\infty = \ & \inf & & \int h_T(T,\xvec) d \mu_T + \int h(t,\xvec) d \mu & \\
  & \ \text{s.t.} & & \left(\frac{\partial \mu}{\partial t} + \div \fvec_1 \mu \right) + \mu_T = \mu_0 & \\
  & & & \left(\frac{\partial \nu}{\partial t} + \div \fvec_2 \nu \right) + \nu_T = \nu_0 & \\
  & & & {\mpi[t,\fyvec \mpnd ] \mu = \mpi[t,\fyvec \mpnd] \nu} & \\
    & & & \int \mu_0 = 1, \quad \int \nu_0 = 1,
  \end{aligned}
  \label{eqn:acc_pmeas}
\end{equation}
{with marginal $\mpi[t,\fyvec \mpnd ] \mu$ respectively $\mpi[t,\fyvec \mpnd ] \nu$ of measure $\mu$ respectively $\nu$ with respect to variables $t$, $\fyvec$.} This form was first derived in \cite{ref:matteo} and is useful for applications where there are a large number of states in \cref{eqn:acc_polynomial} (the new maximum problem size is $1 + n + m$ variables versus $1 + 2n$). The main advantage of this approach is that the complexity of the problem depends on the size of the largest SDP block. This effectively allows us to use our framework to solve much larger problems. 

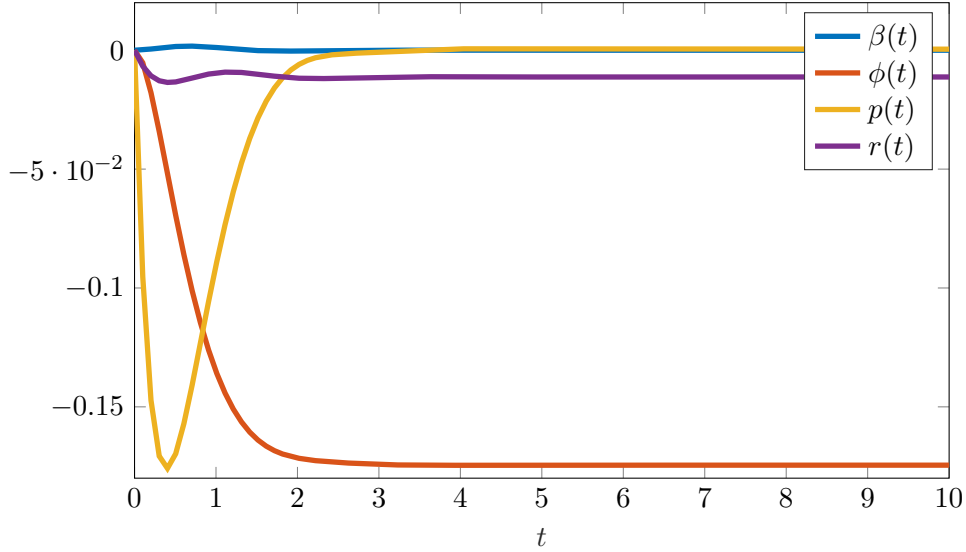
\begin{figure}[t]
  \centering
%
%
\definecolor{mycolor1}{rgb}{0.00000,0.44700,0.74100}%
\definecolor{mycolor2}{rgb}{0.85000,0.32500,0.09800}%
\definecolor{mycolor3}{rgb}{0.92900,0.69400,0.12500}%
\definecolor{mycolor4}{rgb}{0.49400,0.18400,0.55600}%
\begin{tikzpicture}

\begin{axis}[%
width=0.951\figurewidth,
height=\figureheight,
at={(0\figurewidth,0\figureheight)},
scale only axis,
xmin=0,
xmax=10,
xlabel style={font=\color{white!15!black}},
xlabel={$t$},
ymin=-0.18,
ymax=0.02,
axis background/.style={fill=white},
legend style={legend cell align=left, align=left, draw=white!15!black}
]
\addplot [color=mycolor1, line width=2.0pt]
  table[row sep=crcr]{%
0	-0\\
0.202020202020202	0.000510380593993887\\
0.505050505050505	0.0015743438397493\\
0.707070707070708	0.00172579367944969\\
1.01010101010101	0.00113295924347057\\
1.51515151515152	-0.000159787224715657\\
1.91919191919192	-0.00035240369674483\\
3.33333333333333	3.30955909664254e-05\\
5.35353535353535	3.80304695823952e-06\\
10	1.45291245701173e-08\\
};
\addlegendentry{$\beta(t)$}

\addplot [color=mycolor2, line width=2.0pt]
  table[row sep=crcr]{%
0	-0\\
0.1010101010101	-0.00527264741311662\\
0.202020202020202	-0.017854626651916\\
0.303030303030303	-0.0341515813724698\\
0.505050505050505	-0.0694084205261483\\
0.606060606060606	-0.0860041421978135\\
0.707070707070708	-0.101116572750598\\
0.808080808080808	-0.114507646629487\\
0.909090909090908	-0.126109773845517\\
1.01010101010101	-0.135965154561084\\
1.11111111111111	-0.144184633604743\\
1.21212121212121	-0.150919589320104\\
1.31313131313131	-0.156342465080112\\
1.41414141414141	-0.160633065709503\\
1.51515151515152	-0.163968792192545\\
1.61616161616162	-0.166517689291524\\
1.71717171717172	-0.168433624485155\\
1.81818181818182	-0.169853177748342\\
2.02020202020202	-0.171654134756274\\
2.22222222222222	-0.172632042853488\\
2.62626262626263	-0.173643499462711\\
3.23232323232323	-0.174441462933427\\
3.93939393939394	-0.174583002064752\\
6.66666666666667	-0.174530419545102\\
10	-0.174531870063186\\
};
\addlegendentry{$\phi(t)$}

\addplot [color=mycolor3, line width=2.0pt]
  table[row sep=crcr]{%
0	-0\\
0.1010101010101	-0.0953311512727346\\
0.202020202020202	-0.147266176845726\\
0.303030303030303	-0.170696077601299\\
0.404040404040405	-0.175864199336479\\
0.505050505050505	-0.169743379635603\\
0.606060606060606	-0.1570314321827\\
0.707070707070708	-0.140860868298816\\
0.909090909090908	-0.105683405750691\\
1.01010101010101	-0.0888735417335624\\
1.11111111111111	-0.0733897953060723\\
1.21212121212121	-0.0595261374703462\\
1.31313131313131	-0.0474171019305576\\
1.41414141414141	-0.0370839528367632\\
1.51515151515152	-0.0284668065694209\\
1.61616161616162	-0.0214482511249194\\
1.71717171717172	-0.0158718461019447\\
1.81818181818182	-0.0115573878538466\\
1.91919191919192	-0.00831384596270723\\
2.02020202020202	-0.00595029437850059\\
2.12121212121212	-0.00428486581968457\\
2.22222222222222	-0.003151660498947\\
2.42424242424242	-0.00192503138256228\\
2.72727272727273	-0.0012223918363663\\
4.04040404040404	0.000591912909436942\\
10	0.000442021172823104\\
};
\addlegendentry{$p(t)$}

\addplot [color=mycolor4, line width=2.0pt]
  table[row sep=crcr]{%
0	0\\
0.1010101010101	-0.00670263888021694\\
0.202020202020202	-0.0107582140965263\\
0.303030303030303	-0.0128431059839205\\
0.404040404040405	-0.0135389440330762\\
0.505050505050505	-0.0133327141304758\\
0.707070707070708	-0.0116953771834893\\
0.909090909090908	-0.0100240132591196\\
1.11111111111111	-0.00921604253626107\\
1.31313131313131	-0.00934385439930274\\
1.71717171717172	-0.0109102602997879\\
2.02020202020202	-0.0117814808680237\\
2.32323232323232	-0.0119121357191592\\
3.63636363636364	-0.0111483221664166\\
4.94949494949495	-0.011252694472164\\
10	-0.0112472708090259\\
};
\addlegendentry{$r(t)$}

\end{axis}
\end{tikzpicture}%
  \caption{Reference Trajectory $\xvec[r]$ Given $\cvec$} 
  \label{fig:acc_ref}
\end{figure}

For our case, the derivation of the MRAC closed-loop validation problem with sparsity begins by looking at the desired closed-response of \cref{eqn:acc_ref} in \cref{fig:acc_ref}. With the given command signal, we can try approximating the error dynamics using 
\begin{equation}
  \yvec = \begin{bmatrix} \beta(t) - 
\beta_{ss} \\ \phi(t) - \phi_\mathrm{r}(t) \\ p(t) - p_\mathrm{r}(t) \\ r(t) - r_{ss}\end{bmatrix} = \xvec[q] - E \xvec[r] \approx \evec
  \label{eqn:acc_newe}
\end{equation}
where $\beta_{ss} = \lim_{t \rightarrow \infty} \beta_\mathrm{r}(t)$ and $r_{ss} = \lim_{t \rightarrow \infty} r_\mathrm{r}(t)$.

Since $\xvec[r]$ is both linear and \emph{autonomous}, {
we can go directly to \cref{eqn:acc_pmeas} by using \cref{eqn:acc_fx}, piecewise polynomial dynamical optimization \cite{ref:affine} with respect to global occupation measure $\mu$ 
\begin{equation}
  \mu = \mu_1 + \mu_2,
\end{equation}
and the approximated error dynamics \cref{eqn:acc_newe}.} The main takeaway of using \cref{eqn:acc_newe} is that we reduce the total number of equality constraints in \cref{eqn:acc_pmeas} and further simplify the problem. We are now ready to write the MRAC closed-loop problem.


\subsection{MRAC Controller Problem}
\label{sec:acc_mrac}
Given \cref{eqn:acc_pmeas} and the polynomial dynamical optimization problem 
\begin{equation}
  \begin{aligned}
  J = \ & \inf_{\yvec} & & - \norm[\cvec - \tyvec]^2_2 & \\
  & \ \text{s.t.} & &  \dxvec = T D \fvec_j \bigl(t, D^{-1} \xvec, \Lambda_j \uvec \bigl)  &\\
  & & & \dxvec[r] = \refA \xvec[r] + \refB \cvec = \fvec[r] \bigl(\xvec[r], \cvec \bigl)& \\
  & & &  \xvec\in X_j, \quad \xvec[r]  \in X_\mathrm{r}, \quad t \in [0,1], \quad j = 1,2 &\\
  & & & \oxvec \in X_0, \quad \txvec\in X_T, & \\
  & & & \oxvec[r] \in X_{\mathrm{r}0}. \quad \txvec[r] \in X_{\mathrm{r}T} &
  \end{aligned}
  \label{eqn:acc_poly_mrac}
\end{equation}
the new measure-LP for the closed-loop model with MRAC becomes 
\begin{equation}
    \begin{aligned}
  J_\infty = \ & \inf & & - \norm[\cvec - \tyvec]^2_2 & \\
  & \ \text{s.t.} & & \left(\frac{\partial \mu_j}{\partial t} + \div \fvec_j \mu_j \right) + \mu_T = \mu_0, \quad j=1,2 & \\
  & & & \left(\frac{\partial \nu}{\partial t} + \div \fvec[r] \nu \right) + \nu_T = \nu_0 & \\
  & & &  {\mpi[t,\fyvec \mpnd] \mu_1 + \mpi[t,\fyvec \mpnd] \mu_2 = \mpi[t,\fyvec \mpnd] \nu} & \\
    & & & \int \mu_0 = 1, \quad \int \nu_0 = 1,
  \end{aligned}
  \label{eqn:acc_meas_mrac}
\end{equation}
where $\xvec = \begin{bsmallmatrix} \xvec[q], & \wgt, & \xvec[r] \end{bsmallmatrix}$. {We also have marginal $\mpi[t,\fyvec \mpnd ] \mu$ respectively $\mpi[t,\fyvec \mpnd ] \nu$ of measure $\mu$ respectively $\nu$ with respect to variables $t$, $\fyvec$}. With the moment equality constraints and the appropriate supports, the moment LMI relaxation problem of \cref{eqn:acc_meas_mrac} provides a useful upper bound for the cost function $J$ given in \cref{eqn:acc_poly_mrac}.

\begin{figure}[tb]
  \centering \tiny 
  \input{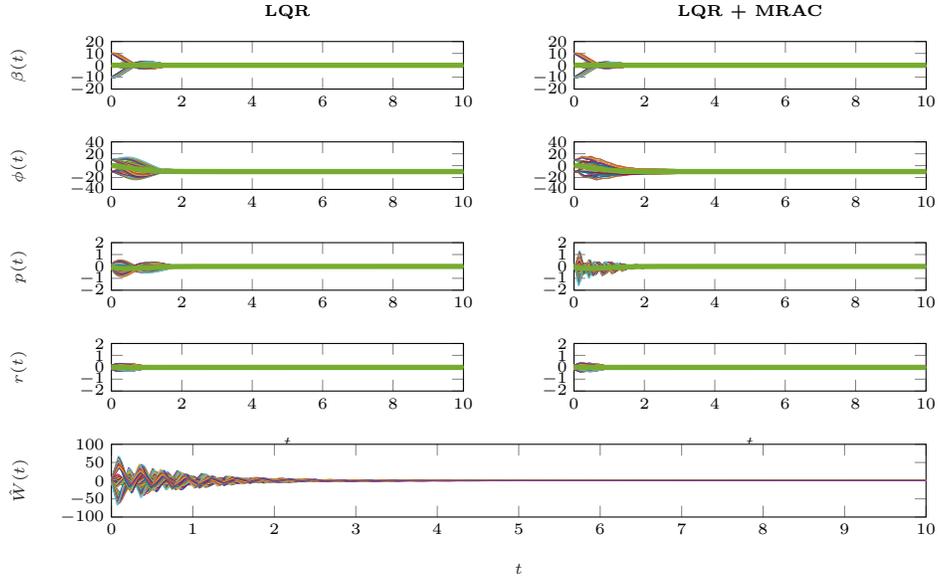} %
  \caption{F-16 Monte-Carlo {with safe trajectories} (${\phi_\mathrm{max} = 1}$)}
  \label{fig:acc_phimax1}
\end{figure}

For both values of $\phi_\mathrm{max}$, good tracking performance was achieved for our LQR + MRAC configuration. This is reflected in our Monte-Carlo simulations \cref{fig:acc_phimax1,fig:acc_phimax2,tab:acc_mc}. Likewise, desirable upper bounds are achieved in our Gloptipoly 3 + MOSEK results \cref{tab:phimax1,tab:phimax2}. The main takeaway is that even a simple LQR + MRAC closed-loop configuration can reject exogenous disturbances and initial condition mismatches. 

It is also possible to achieve improved transient performance using MRAC modifications, such as the error modification \cite{ref:emod} or adaptive loop recovery \cite{ref:alr}, and are discussed extensively in \cite{ref:cdc}. 
\begin{figure}[tb]
  \centering \tiny
  \input{phimax2} %
  \caption{F-16 Monte-Carlo {with unsafe trajectories in baseline LQR} (${\phi_\mathrm{max} = \num{0.314159}}$)}
  \label{fig:acc_phimax2}
\end{figure} %
\begin{table*}[bt]
\caption{GLoptipoly 3 + MOSEK Upper Bounds (${\phi_\mathrm{max} = 1}$)}
\label{tab:phimax1}
\centering
\begin{tabular*}{\textwidth}{@{\extracolsep\fill}lcccc@{\extracolsep\fill}}
\toprule
&\multicolumn{2}{@{}c@{}}{\textbf{LQR}} & \multicolumn{2}{@{}c@{}}{\textbf{LQR + MRAC}} \\ \cmidrule{2-3}\cmidrule{4-5}
\textbf{Rel Ord} & \textbf{Upper Bnd $J$}  & \textbf{CPU [s]}  & \multicolumn{1}{@{}l@{}}{\textbf{Upper Bnd $J$}}  & \textbf{CPU [s]} \\
\midrule
1 & \num{2.5892}	& \num{2.3309} & \num{0.236} & \num{32.36} \\
2 & \num{0.097842}  & \num{2.0548} & \num{0.0006411} & \num{1220.3} \\
3 & \num{0.0014409} & \num{13.307} & \num{1.3964e-05} & \num{2.6475e+05} \\
4 & \num{2.807e-05} & \num{116.72} & -      & - \\
\bottomrule
\end{tabular*}

\end{table*} %
\begin{table*}[bt]
\caption{GLoptipoly 3 + MOSEK Upper Bounds (${\phi_\mathrm{max} = \num{0.314159}}$)}
\label{tab:phimax2}
\centering
\begin{tabular*}{\textwidth}{@{\extracolsep\fill}lcccc@{\extracolsep\fill}}
\toprule
&\multicolumn{2}{@{}c@{}}{\textbf{LQR}} & \multicolumn{2}{@{}c@{}}{\textbf{LQR + MRAC}} \\ \cmidrule{2-3}\cmidrule{4-5}
\textbf{Rel Ord} & \textbf{Upper Bnd $J$}  & \textbf{CPU [s]}  & \multicolumn{1}{@{}l@{}}{\textbf{Upper Bnd $J$}}  & \textbf{CPU [s]} \\
\midrule
1 & \num{2.5892}   & \num{1.4257} & \num{0.22867} 	 & \num{21.956} \\
2 & \num{0.65841}  & \num{1.5347} & \num{0.00064707} & \num{1254.9} \\
3 & \num{0.46795}  & \num{9.6207} & \num{1.5233e-5}  & \num{2.2137e+5} \\
4 & \num{0.45916}  & \num{82.518} & -      	     & - \\
\bottomrule
\end{tabular*}

\end{table*} %
\begin{table}[bt]
\caption{Monte-Carlo Upper Bounds for \Cref{sec:acc_main}}
\label{tab:acc_mc}
\centering
\centering
\begin{tabular*}{\textwidth}{@{\extracolsep\fill}rccc@{\extracolsep\fill}}
\toprule
& {Upper Bound} & {Upper Bound} & \\
$\phi_\mathrm{max}$ & {$J$ (LQR)} & {$J$ (LQR + MRAC)} & {CPU [s]}  \\
\midrule
1 			   & \num{1.8652e-10} & \num{5.0749e-07} & \num{11.7544} \\
\num{0.314159} & \num{0.4458} 	  & \num{5.0749e-07} & \num{11.9667} \\
\bottomrule
\end{tabular*}

\end{table} %


\section{CONCLUSIONS AND FUTURE WORKS}
\label{sec:acc_conc}

We considered a nonlinear dutch-roll F-16 closed-loop model complete with a baseline LQR and MRAC augmentation. To reduce the size of the problem, we exploited sparsity in our framework. These models were then validated using moment LMI relaxations and existing off-the-shelf software. We compared the performance of the closed-loop model baseline LQR controller to the same model with the MRAC. These results were then compared with upper bounds obtained using Monte-Carlo simulations. For future work, we wish to use our framework to consider aircraft models with unknown flexible dynamics.


\end{document}